\documentclass[%
 aip,
 amsmath,amssymb,
 reprint,%
]{revtex4-1}
\usepackage{graphicx,xcolor,hyperref}
\usepackage{braket}
\usepackage[ignoreunlbld,norefs,nocites]{refcheck}
\usepackage{dcolumn}
\usepackage{bm}
\usepackage[utf8]{inputenc}
\usepackage[T1]{fontenc}
\usepackage{mathptmx}
\usepackage{etoolbox}

\makeatletter
\def\@email#1#2{%
 \endgroup
 \patchcmd{\titleblock@produce}
  {\frontmatter@RRAPformat}
  {\frontmatter@RRAPformat{\produce@RRAP{*#1\href{mailto:#2}{#2}}}\frontmatter@RRAPformat}
  {}{}
}%
\makeatother
\begin{document}

\title{Electronic-Entropy-Driven Phase Transitions in Compressed Iron Oxides}
\author{S.\ Azadi}
\altaffiliation[Author to whom correspondence should be addressed: sam.azadi@manchester.ac.uk]
\author{}
\email{sam.azadi@manchester.ac.uk}
\affiliation{Department of Physics and Astronomy, University of Manchester, Oxford Road, Manchester M13 9PL, UK}
\affiliation{Department of Physics, Clarendon Laboratory, University of Oxford, Parks Road, Oxford OX1 3PU, UK}
\author{S.\ M.\ Vinko}
\affiliation{Department of Physics, Clarendon Laboratory, University of Oxford, Parks Road, Oxford OX1 3PU, UK}
\author{C.\ Cr\'{e}pisson}
\affiliation{Department of Physics, Clarendon Laboratory, University of Oxford, Parks Road, Oxford OX1 3PU, UK}
\author{A.\ Principi}
\affiliation{Department of Physics and Astronomy, University of Manchester, Oxford Road, Manchester M13 9PL, UK}
\author{T.\ D.\ K\"{u}hne}
\affiliation{Center for Advanced Systems Understanding, Untermarkt 20, D-02826 G\"orlitz, Germany}
\affiliation{Helmholtz Zentrum Dresden-Rossendorf, Bautzner Landstra{\ss}e 400, D-01328 Dresden, Germany}
\affiliation{TU Dresden, Institute of Artificial Intelligence, Chair of Computational System Sciences, N\"othnitzer Stra{\ss}e 46 D-01187 Dresden, Germany}
\author{M.\ S.\ Bahramy}
\affiliation{Department of Physics and Astronomy, University of Manchester, Oxford Road, Manchester M13 9PL, UK}
\date{\today}

\begin{abstract}
Electronic entropy is usually treated as a secondary correction to structural stability, but under strong electronic excitation it can become a primary thermodynamic driving force. Here we show that electronic entropy can drive both polymorphic and stoichiometric phase transformations in compressed iron oxides. Using finite-temperature density functional theory, we calculate the electronic-temperature-dependent Gibbs free energies of Fe$_2$O, FeO, Fe$_4$O$_5$, Fe$_3$O$_4$, and multiple Fe$_2$O$_3$ polymorphs, including $\alpha$-, $\iota$-, $\zeta$-, $\eta$-, and $\theta$-Fe$_2$O$_3$, over the pressure range 60--260 GPa. At 60-140 GPa, electronic excitation mainly reorganizes the relative stability of Fe$_2$O$_3$ polymorphs, driving transitions from $\iota$-Fe$_2$O$_3$ to $\eta$-Fe$_2$O$_3$. At 180 GPa, the free-energy landscape becomes strongly competitive as FeO is stabilized over an intermediate range of electronic temperature, while $\eta$-Fe$_2$O$_3$ becomes favourable at higher T. At 220-260 GPa, the lowest-free-energy phase at low T is the Fe-rich compound Fe$_2$O, but increasing electronic temperature stabilizes FeO. These results demonstrate that electronic entropy can control not only the relative stability of crystal structures at fixed composition, but also the competition between different iron-oxide stoichiometries. The predicted electronic-entropy-driven phase boundaries provide a route to nonthermal structural transformations in ultrafast and high-energy-density experiments.
\end{abstract}

\maketitle

Iron oxides are among the most important transition-metal compounds in condensed-matter physics, high-pressure science, and planetary materials research. Their structural stability is governed by a complex interplay between Fe-3d states, O-2p bands, charge transfer, magnetic order, spin crossover, and lattice compression. Iron oxides possess several competing stoichiometries and oxidation states, including Fe-rich phases such as Fe$_2$O, rocksalt-like FeO, mixed-valence oxides such as Fe$_4$O$_5$ and Fe$_3$O$_4$, and the multiple polymorphs of Fe$_2$O$_3$ \cite{Cornell2003}. This makes the Fe-O system an ideal platform for studying how electronic degrees of freedom can reshape structural stability under extreme conditions.

The high-pressure behaviour of Fe$_2$O$_3$ is already known to be unusually rich \cite{Cornell2003,Bykova16,Lavina11}. The ambient-pressure $\alpha$-Fe$_2$O$_3$ phase has the corundum structure and remains stable up to several tens of GPa. With increasing pressure, Fe$_2$O$_3$ can transform into the $\iota$-Fe$_2$O$_3$ phase with a Rh$_2$O$_3$(II)-type structure, the $\zeta$-Fe$_2$O$_3$ phase with a distorted perovskite-like structure, and the $\eta$-Fe$_2$O$_3$ phase commonly described as a post-perovskite-type structure. A metastable $\theta$-Fe$_2$O$_3$ phase has also been reported over a restricted pressure-temperature range\cite{Ono2004,Ono2005,Pasternak99,Badro02,Sanson16,Greenberg18}. These structural transformations occur in the same pressure regime where spin transitions, metallization, and changes in Fe-O bonding are expected to play an important role. Recent dynamic-compression experiments have added a further layer of complexity by reporting an isostructural transition from $\alpha$-Fe$_2$O$_3$ to a compressed $\alpha'$-Fe$_2$O$_3$ phase under laser-shock conditions\cite{Amouretti2025}.

Most high-pressure phase transitions in iron oxides have traditionally been interpreted in terms of static enthalpy, lattice temperature, spin crossover, and oxygen fugacity. In this conventional picture, pressure stabilizes denser coordination environments, while lattice temperature enters through vibrational entropy and anharmonic effects. Under strong electronic excitation, however, an additional thermodynamic mechanism becomes possible. If the electronic subsystem reaches an effective temperature of order electronvolts while the ionic lattice remains comparatively cold, electronic entropy can make a large contribution to the free energy. The relevant thermodynamic potential is then not simply the zero-temperature enthalpy, but the finite-temperature electronic Gibbs free energy.

This regime is naturally motivated by ultrafast laser excitation, x-ray free-electron-laser (XFEL) experiments, and high-energy-density conditions\cite{Recoules,Rousse,Medvedev,Sciaini,Amouretti2025,Celin2025,Kang2025,Mazevet,Humphries,Williamson,Silvestrelli,Beaurepaire,Hohlfeld,Zhang0,Carva}. In a two-temperature picture\cite{Kaganov1957,Petrov2021,Allen1987,Carpene2006,Anisimov1975}, electron-electron scattering rapidly establishes a hot electronic distribution, while energy transfer to the lattice occurs on a longer timescale governed by the electron-phonon coupling constant $g$. During this transient window, the ions may remain close to their initial positions, but electronic occupations, screening, bonding, magnetism, and electronic pressure can already be strongly modified. Structural stability can therefore change before conventional lattice heating or melting occurs, providing a route to nonthermal solid-solid transformations driven primarily by the electronic free-energy landscape.

Two complementary experimental routes can access this regime. In XFEL experiments, the x-ray pulse deposits energy directly and preferentially into the electronic subsystem of the sample, driving the electrons to temperatures of order electronvolts on femtosecond timescales while the lattice remains comparatively cold\cite{Vinko12,Kraus25,Humphries}. In laser-driven shock experiments such as those of Cr\'{e}pisson et al.\cite{Celin2025}, the laser couples to an ablator and launches a mechanical shock wave into the sample. At the shock front, rapid compression combined with electronic excitation induced by the passing shock can transiently raise the electronic temperature of the compressed material above that of the ionic lattice, although the magnitude of this electronic heating and its duration depend on the electron-phonon coupling strength and the shock velocity\cite{Petrov2021,Zhang2021,Kang2025}. For transition metals and their oxides, electron-phonon equilibration timescales are typically of order 1--10 ps, which defines the transient window during which the two-temperature description applies\cite{Kaganov1957,Allen1987,Kang2025}. We note that the electronic temperatures considered in this work, spanning 0 to 2.5 eV (approximately 0--29,000 K), represent an upper theoretical envelope intended to map the full electronic free-energy landscape. The temperatures relevant to the predicted phase transitions, for example $T \simeq 0.5$--$0.6$ eV (approximately 6,000--7,000 K) at 140 GPa, are accessible under strong XFEL excitation or at the shock front under high-energy laser drives. Experimental realisation of specific transitions will depend on the achieved electronic temperature, which is controlled by the drive fluence and the electron-phonon coupling of the material at the relevant pressure.

A proper description of this regime requires finite-temperature density functional theory (FT-DFT)\cite{Mermin,Jones2014,Martin2004,Weinert92,Hohenberg,Kresse96,Zhang2021,Driver16,Militzer2021,Wu21,Karasiev16,Karasiev22,SXHu11,Ding18,Bonitz2024}. In Mermin’s extension of DFT, the electronic occupations are determined by the Fermi-Dirac distribution at temperature T, and the electronic entropy is included self-consistently in the free-energy functional. The electronic Helmholtz free energy is $F(V,T)=E(V,T)-T S(V,T),$ where E is the electronic internal energy and S is the electronic entropy. At finite external pressure, the corresponding Gibbs free energy is $G(P,T)=F(V,T)+PV$. This framework allows pressure- and temperature-dependent changes in phase stability to be evaluated directly from the electronic free energy, without assuming that the ionic lattice has reached thermal equilibrium with the electrons.

Magnetic order and local Hubbard interactions are known to play an important role in the electronic structure and phase stability of transition-metal oxides. However, the extreme conditions considered here strongly modify the relevance of both effects. Under high pressure and elevated electronic temperature, the iron oxides studied here become increasingly metallic, enhancing electronic screening and reducing the effective local Hubbard interaction. At the same time, long-range magnetic order is suppressed by electronic excitation. We therefore use standard FT-DFT as a first approximation to isolate the role of electronic entropy in the relative free energies. Since our analysis is based primarily on free-energy differences between competing phases, part of the residual error associated with neglecting explicit magnetic order or a Hubbard correction is expected to cancel. We also note that Hubbard U values commonly used in DFT+U or dynamical mean-field calculations are usually parameterized for near-ground-state electronic configurations. In the present regime, strong electronic excitation can partially depopulate the Fe-d band and substantially modify screening, so the appropriate effective U is itself temperature- and pressure-dependent. Determining this Hubbard-U(P,T) would require a separate study and lies beyond the scope of the present work.

In this work, we use FT-DFT to study electronic-entropy-driven phase stability in compressed iron oxides. We consider hexagonal-Fe$_2$O, B1-FeO, orthorhombic-Fe$_4$O$_5$, orthorhombic-Fe$_3$O$_4$, and several polymorphs of Fe$_2$O$_3$, including $\alpha$-, $\iota$-, $\zeta$-, $\eta$-, and $\theta$-Fe$_2$O$_3$, over the pressure range 60-260 GPa. This expanded set of compounds allows us to distinguish polymorphic transitions within Fe$_2$O$_3$ from changes in the preferred oxide stoichiometry. We find three regimes. At 60-140 GPa, electronic entropy primarily selects between Fe$_2$O$_3$ polymorphs. At 180 GPa, FeO becomes stabilized over an intermediate electronic-temperature window. At 220-260 GPa, Fe$_2$O is the lowest-free-energy phase at low T, but FeO is stabilized as T increases. These results show that electronic entropy can act as an independent thermodynamic control parameter in compressed transition-metal compounds \cite{Azadi2024,Azadi2025,Azadi2025II,AzadiPRM}.

The thermodynamic phase diagrams were obtained using FT-DFT as implemented in the Quantum ESPRESSO package~\cite{QE}. All electronic-structure calculations employed the revised Perdew–Burke–Ernzerhof generalized-gradient approximation for the exchange-correlation functional~\cite{PBE,PBEsol}, together with PAW pseudopotentials supplied with Quantum ESPRESSO~\cite{QE2}. A plane-wave kinetic-energy cutoff of 100 Ry, an augmentation-charge cutoff of 1200 Ry, and a $12\times12\times12$ k-point mesh were used; these parameters were found to give converged total energies and stresses over the full electronic-temperature range considered. Electronic occupations were described by the Fermi–Dirac distribution at electronic temperature T. Since increasing T populates states progressively farther above the Fermi level, the number of empty bands was systematically increased with temperature to ensure convergence of the internal energy, and electronic entropy. For each crystal structure and target pressure, the lattice parameters and atomic coordinates were first optimized at T=50 meV while preserving the crystal symmetry. Subsequent self-consistent FT-DFT calculations were then performed at fixed cell volume for the corresponding pressure. 

To compare compounds with different Fe:O ratios, we calculate a formation Gibbs free energy relative to the same elemental Fe and O atomic references and normalise the result per Fe atom. Specifically, for each phase the formation free energy is
\begin{equation}
G_{\text{form}}(P,T) = \frac{1}{N_{\text{Fe}}}\left[G_{\text{cell}}(P,T) - N_{\text{Fe}}\,\mu_{\text{Fe}} - N_{\text{O}}\,\mu_{\text{O}}\right],
\end{equation}
where $N_{\text{Fe}}$ and $N_{\text{O}}$ are the numbers of Fe and O atoms in the simulation cell, and $\mu_{\text{Fe}}$ and $\mu_{\text{O}}$ are the DFT total energies of the elemental Fe and O references. Dividing by $N_{\text{Fe}}$ yields an intensive quantity (eV per Fe atom) that is well-defined and consistent across all stoichiometries, including Fe$_2$O, FeO, Fe$_4$O$_5$, Fe$_3$O$_4$, and the Fe$_2$O$_3$ polymorphs. The plotted quantity is the relative Gibbs free energy, $\Delta G(P,T)=G_{\text{form},i}(P,T)-G_{\text{form},\alpha\text{-Fe}_2\text{O}_3}(P,T),$ expressed in eV/Fe. This representation places all competing phases on a common formation-free-energy scale and allows the lowest-free-energy candidate to be identified at each $(P,T)$ point. The resulting cross-stoichiometric comparison is defined for the particular Fe and O reference states adopted here. Since phases with different O:Fe ratios respond differently to a change in $\mu_{\rm O}$, the corresponding boundaries should be interpreted as a fixed-reference slice of the broader Fe–O stability space. By contrast, free-energy differences between polymorphs of the same composition are independent of the elemental reference chemical potentials because these terms cancel exactly.

\begin{figure*}
\begin{tabular}{ccc}
    \includegraphics[width=0.33\linewidth]{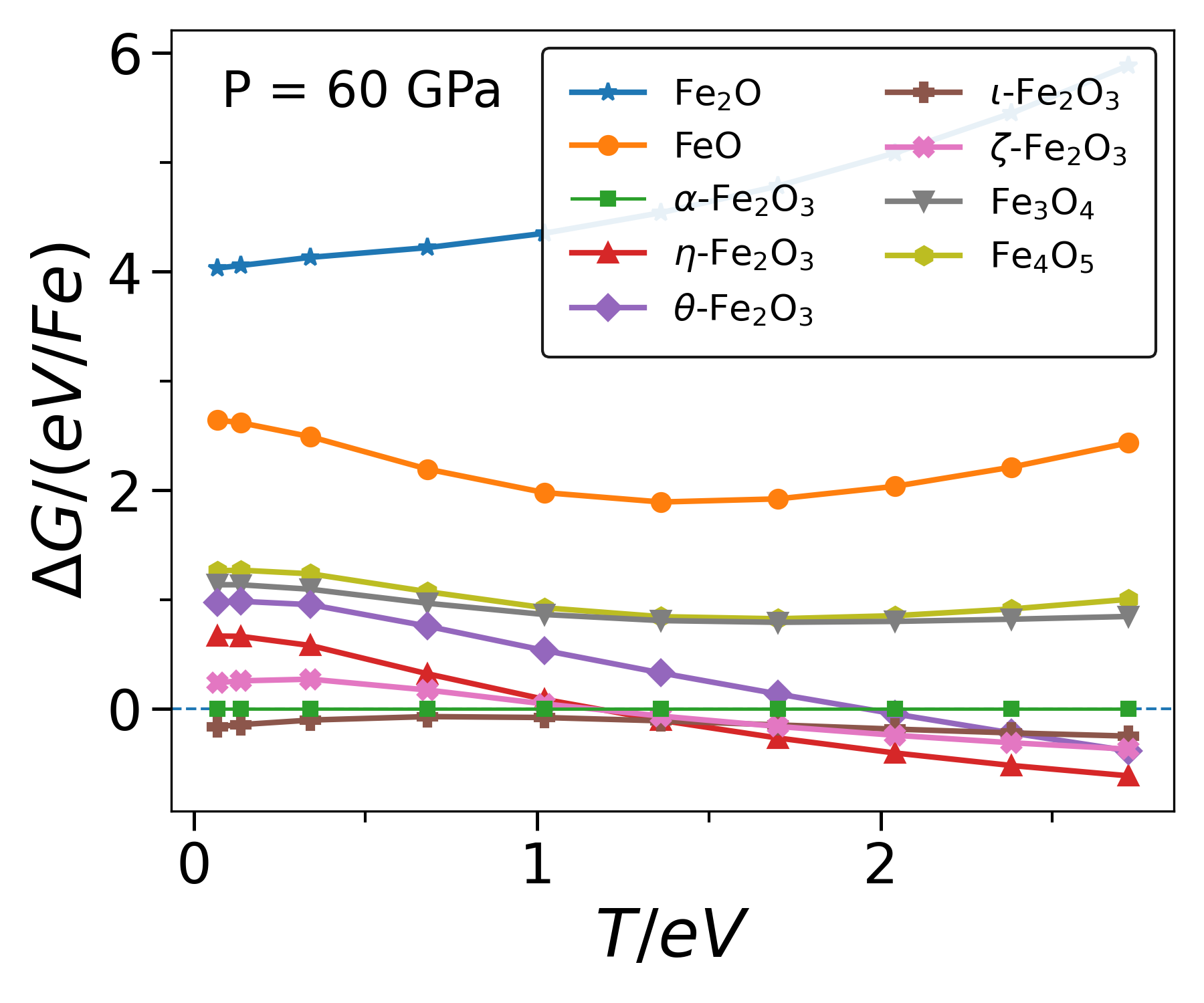}&
    \includegraphics[width=0.33\linewidth]{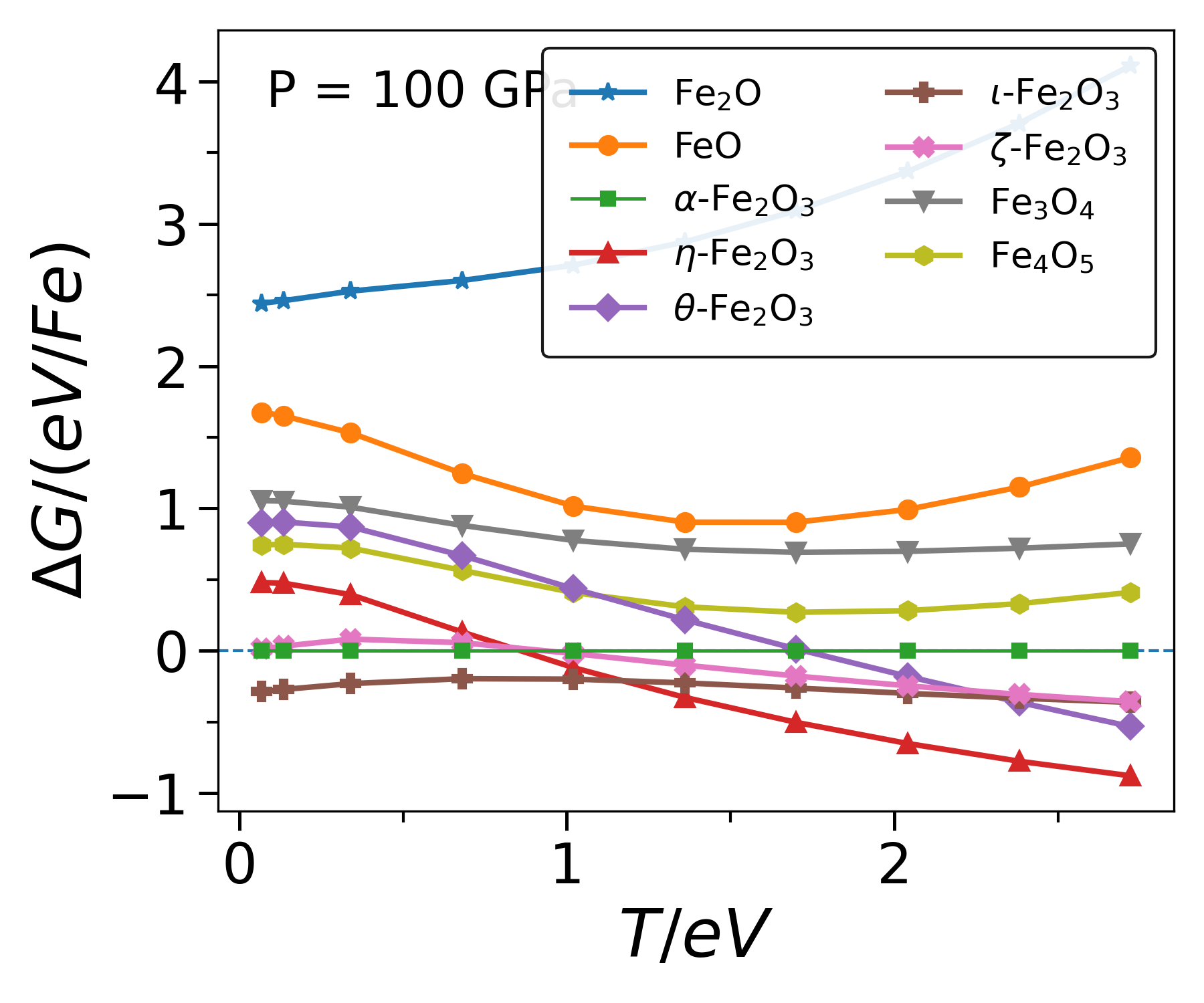}&
    \includegraphics[width=0.33\linewidth]{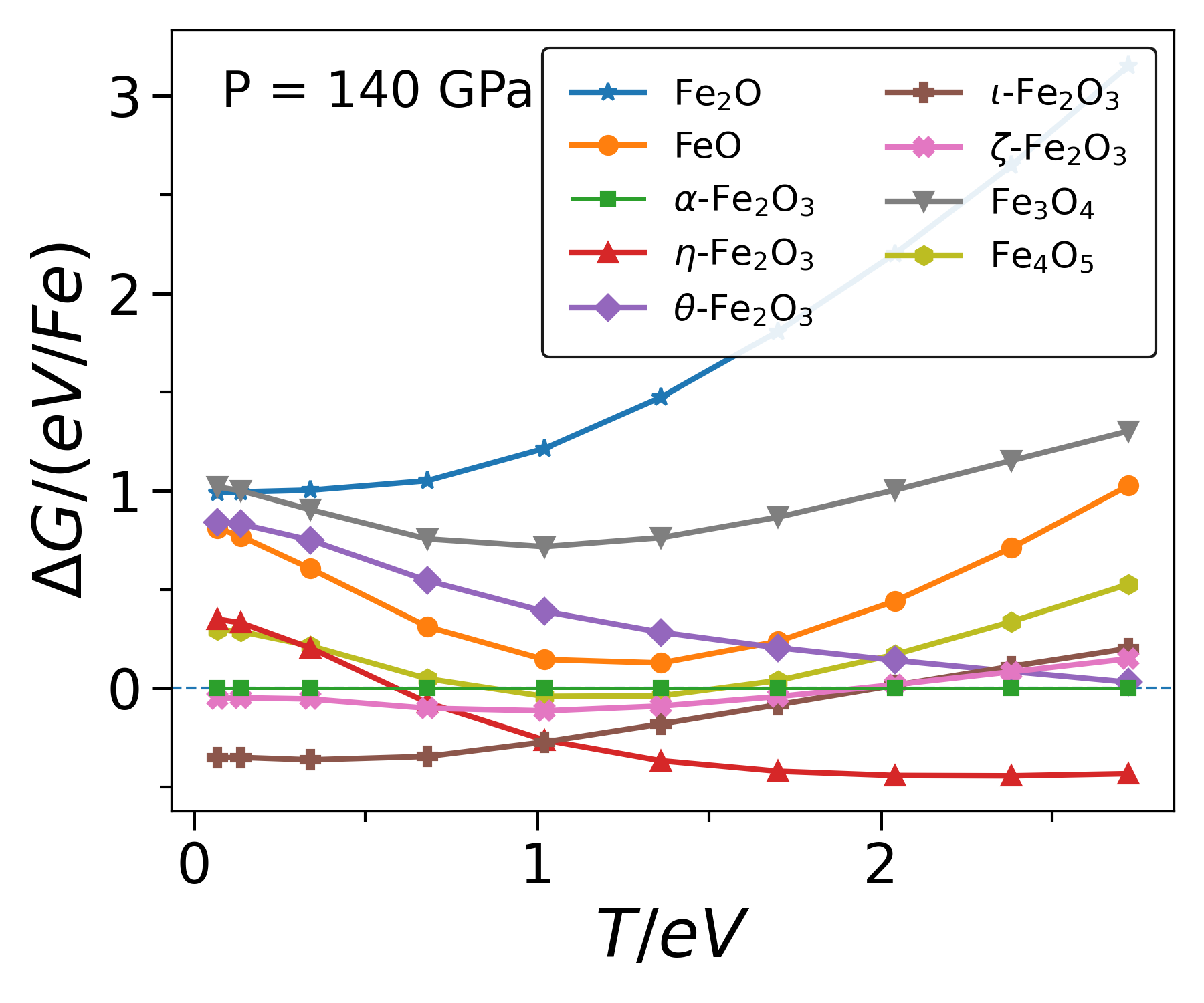}\\
    \includegraphics[width=0.33\linewidth]{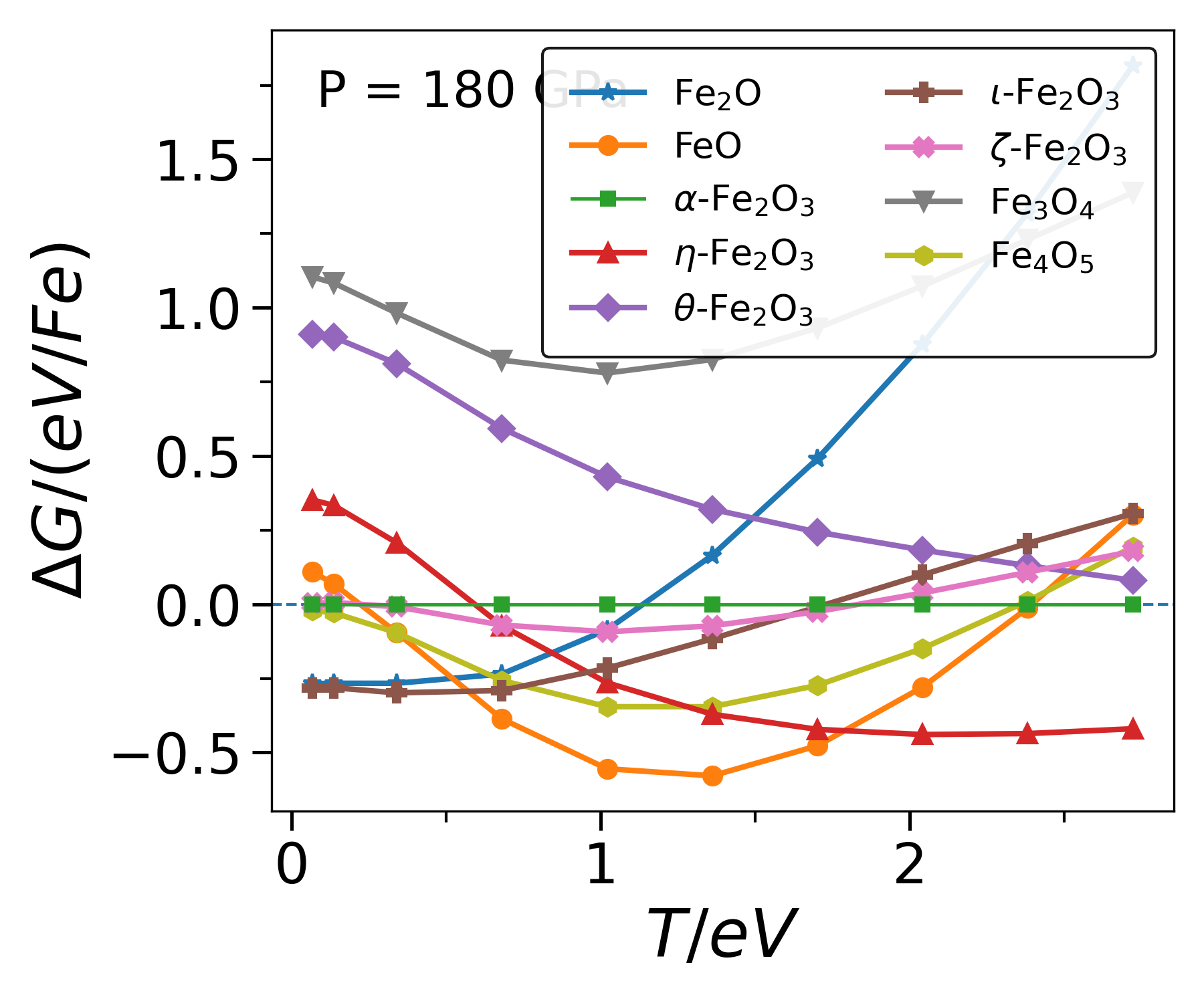}&
    \includegraphics[width=0.33\linewidth]{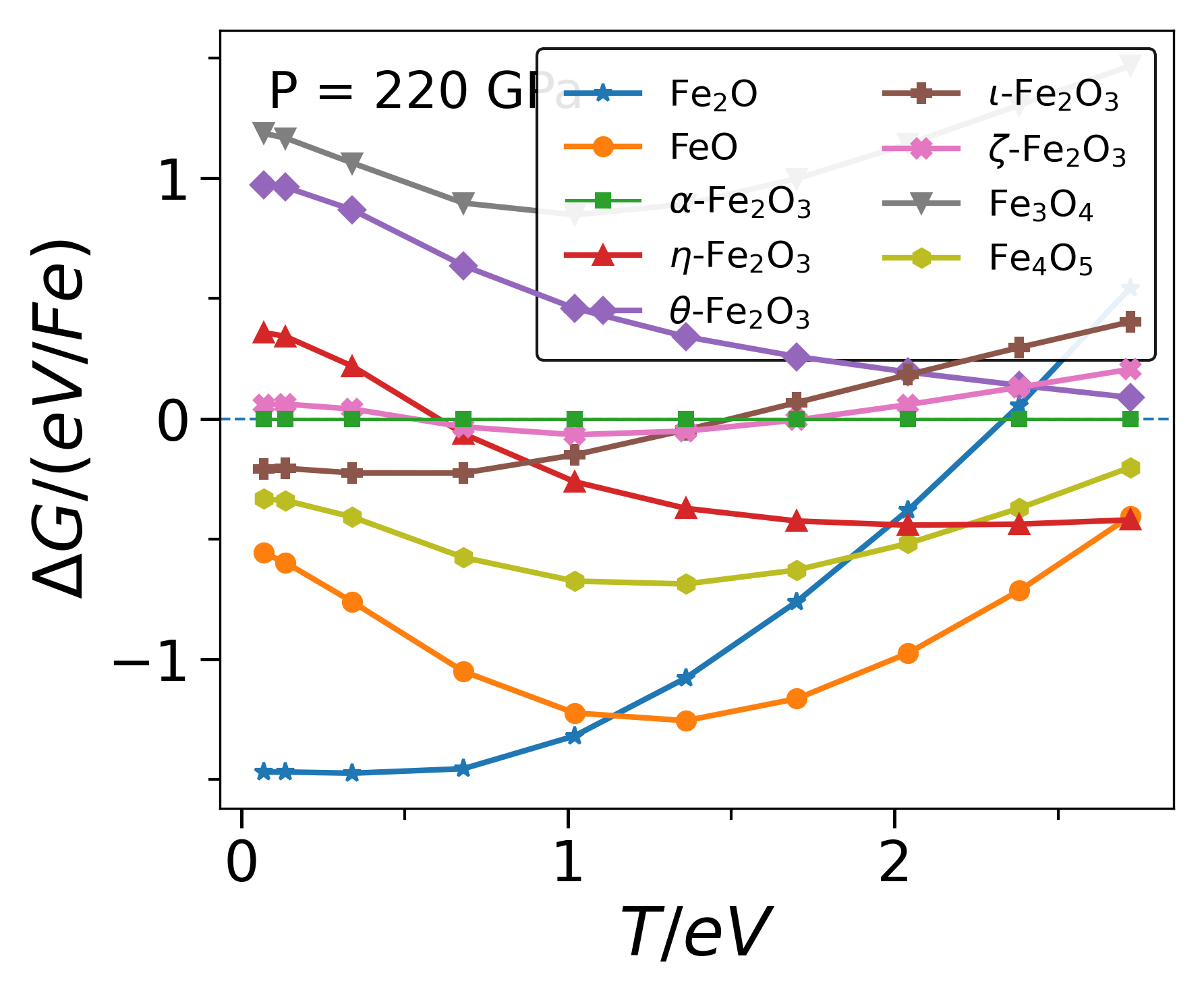}&
    \includegraphics[width=0.33\linewidth]{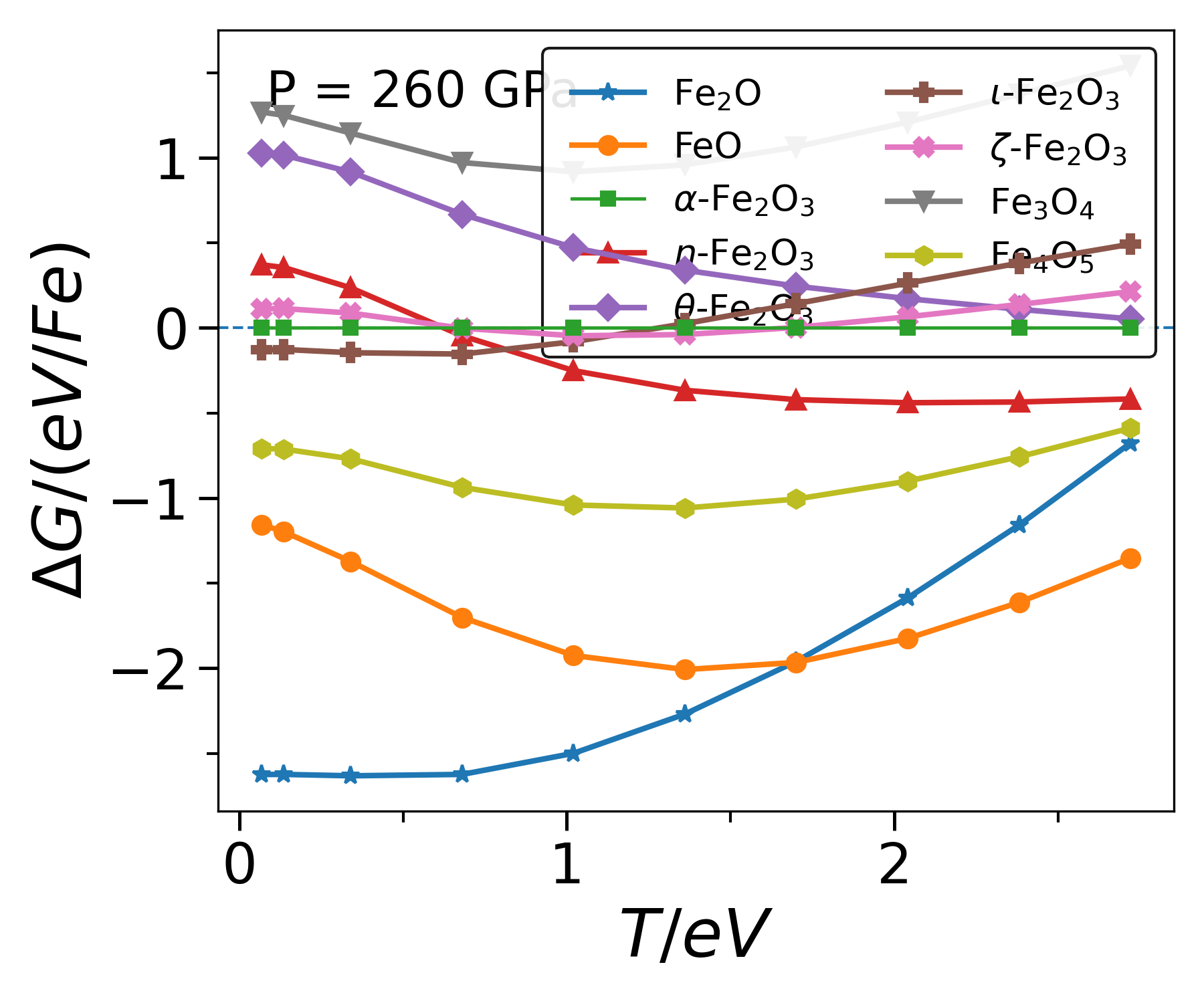}\\
\end{tabular}
\caption{Relative Gibbs free energy $\Delta G$ of Fe$_2$O, FeO, Fe$_4$O$_5$, Fe$_3$O$_4$, and Fe$_2$O$_3$ polymorphs as a function of electronic temperature $T$ at fixed pressures of 60, 100, 140, 180, 220, and 260 GPa. The reference phase is $\alpha$-Fe$_2$O$_3$, so its free energy is set to zero in each panel. All free energies are expressed in eV/Fe (eV per Fe atom) on a common formation-free-energy scale (see text). At 60--140 GPa, the lowest-free-energy phase changes from $\iota$-Fe$_2$O$_3$ to $\eta$-Fe$_2$O$_3$ with increasing $T$, showing an electronic-entropy-driven polymorphic transition within Fe$_2$O$_3$. At 180 GPa, FeO becomes stabilized over an intermediate temperature window before $\eta$-Fe$_2$O$_3$ becomes favourable at higher $T$. At 220--260 GPa, Fe$_2$O is the lowest-free-energy phase at low $T$, while FeO is stabilized at elevated $T$.}
    \label{fig:phasediagram}
\end{figure*}

\begin{figure}
    \centering
    \includegraphics[width=1.\linewidth]{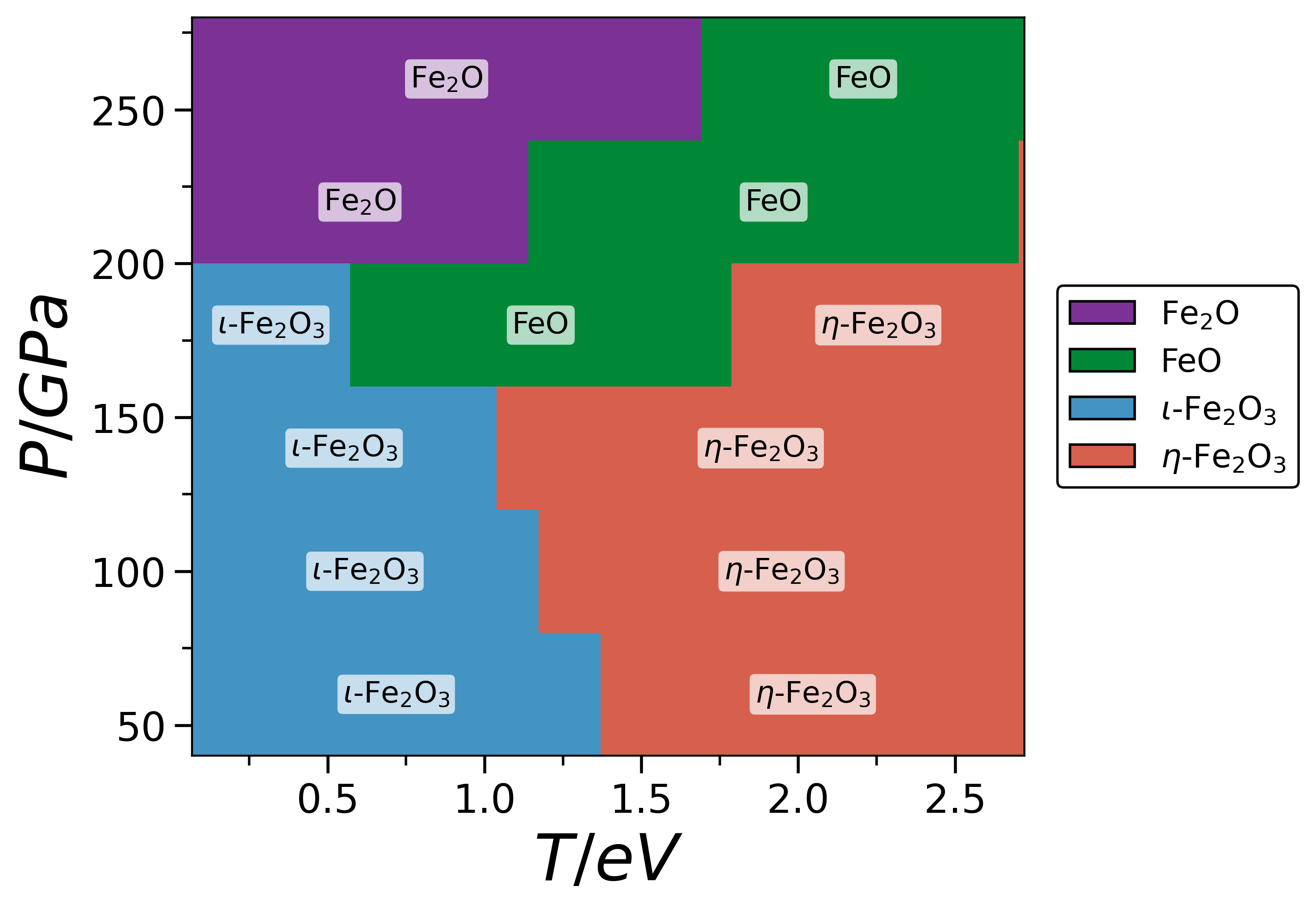}
    \caption{Pressure-temperature phase map showing the lowest Gibbs free-energy phase among the Fe-O structures considered in this work. The horizontal axis is the electronic temperature $T$/eV, and the vertical axis is pressure $P$/GPa. At each $(P,T)$ point, the stable phase is identified by comparing the formation Gibbs free energies of Fe$_2$O, FeO, Fe$_4$O$_5$, Fe$_3$O$_4$, and the $\alpha$-, $\iota$-, $\zeta$-, $\eta$-, and $\theta$-Fe$_2$O$_3$ polymorphs on the same eV/Fe (eV per Fe atom) formation-free-energy scale. The colour of each region denotes the phase with the lowest Gibbs free energy within this candidate set. The map shows a systematic evolution from Fe$_2$O$_3$-polymorph stability at lower pressures to Fe-rich oxide stability at higher pressures. Increasing $T$ drives $\iota$-Fe$_2$O$_3\rightarrow\eta$-Fe$_2$O$_3$ transitions at 60--140 GPa, while at 220--260 GPa electronic excitation stabilizes FeO relative to Fe$_2$O.}
    \label{fig:map}
\end{figure}

It is important to clarify the thermodynamic interpretation of this comparison in the context of the ultrafast regime studied here. Classical equilibrium binary phase diagrams for the Fe-O system are governed by the oxygen chemical potential and require mass transport between phases. In that picture, a change in stoichiometry such as Fe$_2$O$_3 \rightarrow$ Fe$_2$O necessarily involves oxygen release and is controlled by the oxygen fugacity. The present scenario is fundamentally different. In the two-temperature regime, the ionic lattice is effectively frozen on the femtosecond-to-picosecond timescale of electronic excitation: there is no atomic diffusion, no mass transport between phases, and no change in composition. The formation free energies computed here are therefore evaluated at fixed ionic configurations for each candidate phase. The free-energy crossings in Figs.~\ref{fig:phasediagram} and~\ref{fig:map} do not represent diffusional decomposition reactions; rather, they indicate which pre-existing compressed phase is thermodynamically favoured by the electronic free energy at a given $(P,T)$ point. This is the standard convex-hull construction used in computational materials science to identify stable phases\cite{Grimvall}, applied here to the electronic free-energy landscape rather than the zero-temperature enthalpy. The stoichiometric transitions predicted at 220--260 GPa correspond to a change in which bulk phase would nucleate from the excited solid if the electronic temperature were the thermodynamic control parameter, not to a chemical reaction requiring atomic rearrangement on ultrafast timescales. 

Figure \ref{fig:phasediagram} shows the relative Gibbs free energy of Fe$_2$O, FeO, Fe$_4$O$_5$, Fe$_3$O$_4$, and the $\alpha$-, $\iota$-, $\zeta$-, $\eta$-, and $\theta$-Fe$_2$O$_3$ polymorphs as a function of electronic temperature T at fixed pressures between 60 and 260 GPa. The free energies are plotted relative to $\alpha$-Fe$_2$O$_3$, which is therefore set to zero in each panel. Crossings between curves identify changes in the lowest-free-energy phase driven by electronic excitation at fixed pressure. At 60 GPa, the low-T free-energy landscape is dominated by Fe$_2$O$_3$ polymorphs. The $\iota$-Fe$_2$O$_3$ phase lies slightly below the $\alpha$-Fe$_2$O$_3$ reference and is the lowest-free-energy structure at low electronic temperature. With increasing T, the free energy of $\eta$-Fe$_2$O$_3$ decreases rapidly and eventually becomes lower than that of $\iota$-Fe$_2$O$_3$. Thus, the first entropy-driven transition in this pressure range is a polymorphic transformation within Fe$_2$O$_3$, $\iota\text{-Fe}_2\text{O}_3 \rightarrow \eta\text{-Fe}_2\text{O}_3$. Fe-rich phases do not compete at this pressure as Fe$_2$O and FeO remain several eV/Fe above the reference, while Fe$_4$O$_5$ and Fe$_3$O$_4$ remain positive in relative free energy over the full temperature range. 

At 100 GPa, the same qualitative behaviour persists, but the $\iota$-to-$\eta$ transition shifts to lower T. The $\iota$-Fe$_2$O$_3$ phase remains the lowest-free-energy phase at low temperature, while $\eta$-Fe$_2$O$_3$ is increasingly stabilized as T rises. The crossing occurs near T$\simeq 1.1$ eV. The $\theta$- and $\zeta$-Fe$_2$O$_3$ phases also decrease in free energy with increasing T, but they do not become the lowest-free-energy structures. FeO, Fe$_2$O, Fe$_3$O$_4$, and Fe$_4$O$_5$ remain above the stable Fe$_2$O$_3$ polymorphs. 

At 140 GPa, the electronic-entropy-driven $\iota$-Fe$_2$O$_3$$\rightarrow\eta$-Fe$_2$O$_3$ transition moves to still lower T, occurring around T$\simeq$ 0.5-0.6 eV. The $\eta$ phase then remains the lowest-free-energy phase over the rest of the plotted temperature range. This pressure therefore marks the strongest region of Fe$_2$O$_3$-polymorph selection as the electronic excitation reorganizes the relative stability of Fe$_2$O$_3$ phases without stabilizing FeO or Fe-rich oxides. Fe$_4$O$_5$ becomes much closer in free energy than at 60 and 100 GPa, and even approaches the $\alpha$-Fe$_2$O$_3$ reference near intermediate T, but it remains above $\eta$-Fe$_2$O$_3$. This shows that intermediate Fe-rich oxides are becoming thermodynamically relevant, although they do not yet control the phase boundary.

A qualitatively different regime appears at 180 GPa. At low T, several phases become nearly competitive: $\iota$-Fe$_2$O$_3$, Fe$_2$O, FeO, and Fe$_4$O$_5$ all lie close to the $\alpha$-Fe$_2$O$_3$ reference. As T increases, FeO is strongly stabilized and becomes the lowest-free-energy phase over an intermediate electronic-temperature window. At higher T, however, the FeO curve turns upward, while $\eta$-Fe$_2$O$_3$ continues to decrease and becomes the lowest-free-energy phase. The resulting sequence is therefore approximately
$\iota\text{-Fe}_2\text{O}_3\rightarrow\mathrm{FeO}\rightarrow\eta\text{-Fe}_2\text{O}_3$. This pressure is the crossover regime between Fe$_2$O$_3$-dominated polymorphic transitions and Fe-rich oxide stabilization. 

At 220 GPa, the lowest-free-energy phase at low T is Fe$_2$O. This Fe-rich phase lies far below the $\alpha$-Fe$_2$O$_3$ reference and remains strongly stabilized up to intermediate electronic temperature. As T increases, the free energy of FeO decreases more rapidly and crosses that of Fe$_2$O, making FeO the lowest-free-energy phase at higher T. Thus, at 220 GPa the dominant entropy-driven transformation is a stoichiometric transition between two Fe-rich oxides, $\mathrm{Fe}_2\mathrm{O}\rightarrow\mathrm{FeO}$. The Fe$_4$O$_5$ phase is also stabilized relative to $\alpha$-Fe$_2$O$_3$, but it remains above both Fe$_2$O and FeO. The Fe$_2$O$_3$ polymorphs are no longer the lowest-free-energy phases, although $\eta$-Fe$_2$O$_3$ remains the most favourable among them at elevated T.

At 260 GPa, the Fe-rich character of the low-temperature phase diagram becomes even more pronounced. Fe$_2$O is the most stable phase at low T, lying well below all Fe$_2$O$_3$ polymorphs and below FeO. Increasing electronic temperature again stabilizes FeO, which becomes lower in free energy than Fe$_2$O at intermediate-to-high T. The sequence $\mathrm{Fe}_2\mathrm{O}\rightarrow\mathrm{FeO}$ therefore persists at the highest pressure considered. Fe$_4$O$_5$ remains  significantly lower than the Fe$_2$O$_3$ polymorphs over much of the temperature range, but it does not become the lowest-free-energy phase. This result shows that the high-pressure phase stability is governed by competition between Fe-rich oxides rather than by Fe$_2$O$_3$ polymorphism.

The pressure evolution of the phase diagram can be summarized in three regimes. In the first regime, 60-140 GPa, the lowest-free-energy phases are Fe$_2$O$_3$ polymorphs, and electronic entropy drives $\iota$-Fe$_2$O$_3\rightarrow\eta$-Fe$_2$O$_3$ transitions. In the second regime, around 180 GPa, the system enters a crossover region where FeO is stabilized over an intermediate temperature interval, while $\eta$-Fe$_2$O$_3$ becomes favourable at higher T. In the third regime, 220-260 GPa, Fe-rich oxides dominate: Fe$_2$O is favoured at low T, whereas FeO is stabilized by electronic entropy at higher T. The overall pressure-temperature evolution is therefore
$\iota\text{-Fe}_2\text{O}_3\rightarrow\eta\text{-Fe}_2\text{O}_3\rightarrow\mathrm{Fe}_2\mathrm{O}\rightarrow\mathrm{FeO}$,
with the exact sequence depending on pressure and temperature.

These results demonstrate that electronic entropy can control phase stability in two distinct ways. At fixed stoichiometry, it can reorder competing crystal structures, as seen in the $\iota$-Fe$_2$O$_3\rightarrow\eta$-Fe$_2$O$_3$ transitions. Across different stoichiometries, it can alter the balance between Fe-rich and O-rich compounds, as seen in the stabilization of Fe$_2$O and FeO at high pressure. This second effect is particularly important because it shows that electronic excitation can move the system across a compositional free-energy landscape, not only a polymorphic one.

The slopes and curvatures of the $\Delta \text{G(T)}$ curves provide direct evidence for the role of electronic entropy. A phase whose Gibbs free energy decreases more rapidly with T gains more electronic free-energy stabilization relative to the reference. The rapid stabilization of $\eta$-Fe$_2$O$_3$ at 60-140 GPa indicates a larger entropy contribution for this phase relative to $\iota$- and $\alpha$-Fe$_2$O$_3$. At higher pressures, the strong nonmonotonic behaviour of FeO and Fe$_2$O reflects a competition between electronic entropy, internal electronic energy, and the pressure-volume term. The resulting free-energy crossings are therefore not small perturbative shifts, but represent changes in the thermodynamic ground state within the set of phases considered.

The summary phase map in Fig.~\ref{fig:map} condenses the free-energy crossings shown in Fig.~\ref{fig:phasediagram} into a single pressure-temperature diagram. Three distinct regimes are obtained. At 60–140 GPa, the lowest-free-energy phases are Fe$_2$O$_3$ polymorphs: $\iota$-Fe$_2$O$_3$ is favoured at low T, whereas increasing electronic temperature stabilizes $\eta$-Fe$_2$O$_3$. This region therefore corresponds to an electronic-entropy-driven polymorphic transition at fixed composition. Around 180 GPa, the phase diagram enters a crossover regime in which FeO competes with both Fe$_2$O$_3$ and Fe-rich oxides, indicating that electronic entropy begins to alter the balance between different Fe:O ratios. At 220–260 GPa, Fe-rich phases dominate the low-temperature free-energy landscape as Fe$_2$O is the lowest-free-energy phase at low $T$, but FeO becomes stabilized as $T$ increases. The map therefore shows that electronic entropy changes the phase stability of compressed iron oxides in two ways: it first selects between competing Fe$_2$O$_3$ polymorphs, and at higher pressure it drives transitions between different oxide stoichiometries. 

The present results connect directly to recent experimental and theoretical work on Fe$_2$O$_3$ under laser-driven shock compression. Cr\'{e}pisson et al. \cite{Celin2025,Celin26} reported time-resolved x-ray diffraction of Fe$_2$O$_3$ shocked to pressures between 122 and 209 GPa, observing the onset of diffuse scattering at 122–131 GPa and attributing a rapid change in the structure factor between 145 and 151 GPa to an amorphous-to-liquid transition. To interpret those experiments, Hugoniot temperatures and elastic moduli were computed using DFT+U, with a Hubbard correction parameterised to reproduce the known ambient band gap and magnetic moment of $\alpha$-Fe$_2$O$_3$. That choice was appropriate for a study in which the ionic temperature is the thermodynamic variable of interest and the electronic subsystem is assumed to remain in its ground state. In the present work the focus is precisely on what happens when this assumption breaks down meaning under ultrafast excitation the electronic subsystem can reach temperatures of order electronvolts before the lattice equilibrates, and it is the electronic entropy, which is absent from a ground-state DFT+U description, that reshapes the free-energy landscape. Standard FT-DFT without a Hubbard correction is the natural framework for this regime, because strong electronic excitation progressively delocalises the Fe-3d states and suppresses the local-moment physics that the Hubbard U is designed to capture. The two computational approaches are therefore complementary: DFT+U describes the cold-lattice Hugoniot accurately, while FT-DFT captures the transient free-energy landscape during the window of elevated electronic temperature that precedes lattice equilibration.

The pressure range in which Cr\'{e}pisson et al. observe shock amorphisation, 122–145 GPa, falls within the regime identified here as dominated by electronic-entropy-driven competition among Fe$_2$O$_3$ polymorphs. Our calculations show that in this pressure range, increasing electronic temperature strongly stabilises $\eta$-Fe$_2$O$_3$ relative to the $\iota$-Fe$_2$O$_3$ phase present at low temperature, lowering the free-energy crossing to T $\simeq$ 0.5–0.6 eV at 140 GPa. Cr\'{e}pisson et al. further demonstrate, through DFT+U elastic-constant calculations, that the shear and Young's moduli of $\alpha'$-Fe$_2$O$_3$ become negative at 120 GPa, indicating that the compressed crystal cannot sustain the applied shear stress. Taken together, these two results suggest the following picture: the transient electronic free-energy landscape already favours structural reorganisation toward $\eta$-Fe$_2$O$_3$, but the polymorph cannot nucleate on the timescale of the shock; the mechanically unstable $\alpha'$-Fe$_2$O$_3$ therefore collapses into an amorphous phase instead, with the electronic entropy providing the thermodynamic driving force and the negative elastic moduli providing the kinetic pathway. The melt onset observed experimentally above 151 GPa is broadly consistent with the crossover behaviour we predict near 180 GPa, where the free-energy landscape becomes highly competitive among Fe$_2$O$_3$ polymorphs, FeO, and Fe-rich oxides simultaneously. Hence, the quantitative difference in pressure likely reflects the distinct thermodynamic path sampled along the shock Hugoniot relative to the fixed-pressure, variable-electronic-temperature trajectories computed here. Extending these experiments to 220–260 GPa, where our calculations predict an electronic-entropy-driven stoichiometric transition from Fe$_2$O to FeO, would provide a direct test of the phase boundaries reported in Fig.~\ref{fig:map}.

The phase boundaries reported here identify the thermodynamic conditions under which electronic entropy favors one phase over another, but they do not determine whether, or how rapidly, the corresponding transformation will occur. Structural phase transformations require the nucleation and growth of a product phase, processes that depend on interfacial energies, activation barriers, defects, and the ability of the lattice to undergo the necessary atomic rearrangements. These kinetic effects lie beyond the scope of the present FT-DFT free-energy calculations, which describe only the relative thermodynamic stability of competing phases under electronic excitation. A free-energy crossing should therefore be interpreted as identifying the thermodynamic driving force for a transformation rather than predicting its pathway or rate. Quantitative prediction of transformation kinetics will require multiscale approaches that combine the electronic free-energy landscape obtained from FT-DFT with explicit models of nucleation and growth. Although ab initio molecular dynamics can capture the earliest stages of atomic rearrangement, its accessible timescales are generally too short to observe nucleation directly. Enhanced-sampling techniques, including umbrella sampling and metadynamics, can determine nucleation barriers and rates, while phase-field models can describe the subsequent evolution of microstructure over experimentally relevant length and timescales \cite{Bussi20}. An important future direction will therefore be the development of multiscale frameworks in which FT-DFT provides the pressure- and electronic-temperature-dependent free-energy landscape, while atomistic and mesoscale models describe the nucleation, growth, and propagation of the resulting phases.

In summary, FT-DFT calculations reveal a systematic pressure-dependent evolution of electronic-entropy-driven phase stability in compressed iron oxides. At 60-140 GPa, electronic entropy drives polymorphic transitions within Fe$_2$O$_3$, selecting $\eta$-Fe$_2$O$_3$ at elevated T. At 180 GPa, FeO becomes stabilized over an intermediate electronic-temperature window, marking the onset of stoichiometric competition. At 220-260 GPa, Fe$_2$O is the preferred low-T phase, but electronic excitation stabilizes FeO. These results show that electronic entropy is an independent thermodynamic driving force capable of reshaping both structure and stoichiometry in transition-metal compounds under compression.

\textit{Acknowledgment.} S.A, A.P, and M.S.B acknowledge the support of the Leverhulme Trust under the grant agreement RPG-2023-253. S. Azadi and T.D. K\"{u}hne acknowledge the computing time provided to them on the high-performance computers Noctua2 at the NHR Center in Paderborn (PC2). S.M.V. acknowledges support from the UK EPSRC under grant EP/W010097/1. The data that support the findings of this article are openly available \cite{github}, embargo periods may apply.

\textit{Conflict of interest.} The authors have no conflicts to disclose.

\bibliography{main}

@article{Grimvall,
  title={Lattice instabilities in metallic elements},
  author={G. Grimwall and B. Magyari-K\"{o}pe and V. Ozolin\v{s} and K. A. Persson},
  journal={Rev. Mod. Phys.},
  year={2012},
  volume={84},
  pages={945}
}

@book{Martin2004,
  title={Electronic Structure: Basic Theory and Practical Methods},
  author={R.M. Martin},
  publisher = {Cambridge University Press},
  year={2004}
}

@article{Recoules,
  title={Effect of intense laser irradiation on the lattice stability of semiconductors and metals},
  author={V. Recoules and J. Cl\'{e}rouin and G. Z\'{e}rah and P.M. Anglade and S. Mazevet},
  journal={Phys. Rev. Lett.},
  year={2006},
  volume={96},
  pages={055503}
}

@article{Rousse,
  title={Non-thermal melting in semiconductors measured at femtosecond resolution},
  author={A. Rousse and C. Rischel and S. Fourmaux and I. Uschmann and S. Sebban and et al.},
  journal={Nature (London)},
  year={2001},
  volume={410},
  pages={65}
}

@article{Medvedev,
  title={Nonthermal phase transitions in metals},
  author={N. Medvedev and I. Milov},
  journal={Scientific Reports},
  year={2020},
  volume={10},
  pages={12775}
}

@article{Sciaini,
  title={Femtosecond electron diffraction: heralding the era of atomically resolved dynamics},
  author={G. Sciaini and R. J. D. Miller},
  journal={Rep. Prog. Phys.},
  year={2011},
  volume={74},
  pages={096101}
}

@article{QE,
  title={QUANTUM ESPRESSO: a modular and open-source software project for quantum simulations of materials},
  author={P. Giannozzi and S. Baroni and N. Bonini and M. Calandra and R. Car and et al.},
  journal={J. Phys.: Condens. Matter},
  year={2009},
  volume={21},
  pages={395502}
}

@article{Hohenberg,
  title={Inhomogeneous Electron Gas},
  author={P. Hohenberg and W. Kohn},
  journal={Phys. Rev.},
  year={1964},
  volume={136},
  pages={B864}
}

@article{PBE,
  title={Generalized Gradient Approximation Made Simple},
  author={John P. Perdew and Kieron Burke and Matthias Ernzerhof},
  journal={Phys. Rev. Lett.},
  year={1996},
  volume={77},
  pages={3865}
}

@article{PBEsol,
  title={Restoring the Density-Gradient Expansion for Exchange in Solids and Surfaces},
  author={J. P. Perdew and A. Ruzsinszky and G. I. Csonka and O. A. Vydrov and G. E. Scuseria and et al.},
  journal={Phys. Rev. Lett.},
  year={2008},
  volume={100},
  pages={136406}
}

@article{QE2,
  title={Advanced capabilities for materials modelling with Quantum ESPRESSO},
  author={P. Giannozzi and O. Andreussi and T. Brumme and O. Bunau and M. Buongiorno Nardelli and et al.},
  journal={J. Phys.: Condens. Matter},
  year={2017},
  volume={29},
  pages={465901}
}

@article{Mazevet,
  title={Ab-Initio Simulations of the Optical Properties of Warm Dense Gold},
  author={S. Mazevet and J. Cl\'{e}rouin and V. Recoules and P. M. Anglade and G. Zerah},
  journal={Phys. Rev. Lett.},
  year={2005},
  volume={95},
  pages={085002}
}

@article{Humphries,
  title={Probing the Electronic Structure of Warm Dense Nickel via Resonant Inelastic X-Ray Scattering},
  author={O. S. Humphries and R. S. Marjoribanks and Q. Y. van den Berg and E. C. Galtier and M. F. Kasim and et al.},
  journal={Phys. Rev. Lett.},
  year={2020},
  volume={125},
  pages={195001}
}

@article{Williamson,
  title={Time-Resolved Laser-Induced Phase Transformation in Aluminum},
  author={S. Williamson and G. Mourou and J. C. M. Li},
  journal={Phys. Rev. Lett.},
  year={1984},
  volume={52},
  pages={2364}
}

@article{Silvestrelli,
  title={Ab initio Molecular Dynamics Simulation of Laser Melting of Silicon},
  author={P. L. Silvestrelli and A. Alavi and M. Parrinello and D. Frenkel},
  journal={Phys. Rev. Lett.},
  year={1996},
  volume={77},
  pages={3149}
}

@article{Beaurepaire,
  title={Ultrafast Spin Dynamics in Ferromagnetic Nickel},
  author={E. Beaurepaire and J.-C. Merle and A. Daunois and J.-Y. Bigot},
  journal={Phys. Rev. Lett},
  year={1996},
  volume={76},
  pages={4250}
}

@article{Hohlfeld,
  title={Nonequilibrium Magnetization Dynamics of Nickel},
  author={J. Hohlfeld and E. Matthias and R. Knorren and K. H. Bennemann},
  journal={Phys. Rev. Lett.},
  year={1997},
  volume={78},
  pages={4861}
}

@article{Zhang0,
  title={Laser-Induced Ultrafast Demagnetization in Ferromagnetic Metals},
  author={G. P. Zhang and W. H\"{u}bner},
  journal={Phys. Rev. Lett.},
  year={2000},
  volume={85},
  pages={3025}
}

@article{Carva,
  title={Ab Initio Investigation of the Elliott-Yafet Electron-Phonon Mechanism in Laser-Induced Ultrafast Demagnetization},
  author={K. Carva and M. Battiato and P. M. Oppeneer},
  journal={Phys. Rev. Lett.},
  year={2011},
  volume={107},
  pages={207201}
}

@article{Mermin,
  title={Thermal properties of the inhomogenous electron gas},
  author={N. D. Mermin},
  journal={Phys. Rev.},
  year={1965},
  volume={137},
  pages={A1441}
}

@article{Jones2014,
  title={Thermal density functional theory in context},
  author={A. Pribram-Jones and S. Pittalis and E.K.U. Gross and K. Burke},
  journal={Frontiers and Challenges in Warm Dense Matter},
  year={2014},
  volume={25},
  pages={60}
}

@article{Azadi2024,
  title={Nonthermal solid-solid phase transition in ferromagnetic iron},
  author={S. Azadi and J. S. Wark and S. M. Vinko},
  journal={Phys. Rev. B},
  year={2024},
  volume={110},
  pages={214434}
}

@article{Azadi2025,
  title={Lattice stability of ultrafast-heated gold},
  author={S. Azadi and J. S. Wark and S. M. Vinko},
  journal={Scientific Reports},
  year={2025},
  volume={15},
  pages={5350}
}

@article{Amouretti2025,
  title={Isostructural Phase Transition of \text{Fe}$_2$\text{O}$_3$ under Laser Shock Compression},
  author={A. Amouretti and C. Crepisson and S. Azadi and F. Brisset and D. Cabaret and et al.},
  journal={Phys. Rev. Lett.},
  year={2025},
  volume={134},
  pages={176102}
}

@article{Celin2025,
  title={Shock-driven amorphization and melting in \text{Fe}$_2$\text{O}$_3$},
  author={C. Crepisson and A. Amouretti and M. Harmand and C. Sanloup and P. Heighway and et al.},
  journal={Phys. Rev. B},
  year={2025},
  volume={111},
  pages={024209}
}

@article{Kang2025,
  title={Nonequilibrium electron-phonon and electron-ion couplings in warm dense copper},
  author={G. Kang and G. Lee and S. Azadi and R. Carley and L. Le Guyader and et al.},
  journal={Applied Surface Science},
  year={2025},
  volume={713},
  pages={164304}
}

@article{Zhang2021,
  title={Finite-temperature density-functional-theory investigation on the nonequilibrium transient warm-dense-matter state created by laser excitation},
  author={Hengyu Zhang and Shen Zhang and Dongdong Kang and Jiayu Dai and M. Bonitz},
  journal={Phys. Rev. E},
  year={2021},
  volume={103},
  pages={013210}
}

@article{Azadi2025II,
  title={Electronic temperature driven phase stability and structural evolution of iron at high pressure},
  author={S. Azadi and S. M. Vinko},
  journal={Phys. Rev. B},
  year={2025},
  volume={112},
  pages={134103}
}

@article{AzadiPRM,
  title={Electronic-entropy-driven solid-solid phase transitions in elemental metals},
  author={S. Azadi and S. M. Vinko and A. Principi and T. D. K\"{u}hne and M. S. Bahramy},
  journal={Phys. Rev. Materials},
  year={2026},
  volume={10},
  pages={045001}
}

@article{Kaganov1957,
  title={Relaxation between electrons and the crystalline lattice},
  author={M.I. Kaganov and I.M. Lifshitz and N.V. Tanatarov},
  journal={Sov. Phys. JETP},
  year={1957},
  volume={4},
  pages={173}
}

@article{Petrov2021,
  title={Modeling of short-pulse laser-metal interactions in the warm dense matter regime using the two-temperature model},
  author={G. M. Petrov and A. Davidson and D. Gordon and J. Penano},
  journal={Phys. Rev. E},
  year={2021},
  volume={103},
  pages={033204}
}

@article{Allen1987,
  title={Theory of thermal relaxation of electrons in metals},
  author={P.B. Allen},
  journal={Phys. Rev. Lett.},
  year={1987},
  volume={59},
  pages={1460}
}

@article{Carpene2006,
  title={Ultrafast laser irradiation of metals: Beyond the two-temperature model},
  author={E. Carpene},
  journal={Phys. Rev. B},
  year={2006},
  volume={74},
  pages={024301}
}

@article{Anisimov1975,
  title={Electron emission from metal surfaces exposed to ultrashort laser pulses},
  author={S. I. Anisimov and B. L. Kapeliovich and T. L. Perelman},
  journal={Sov. Phys. JETP},
  year={1975},
  volume={39},
  pages={375}
}

@article{Weinert92,
  title={Fractional occupations and density-functional energies and forces},
  author={M. Weinert and J. W. Davenport},
  journal={Phys. Rev. B},
  year={1992},
  volume={45},
  pages={13709}
}

@article{Kresse96,
  title={Efficiency of ab-initio total energy calculations for metals and semiconductors},
  author={G. Kresse and J. Furthm\"{u}ller},
  journal={Comp. Mat. Science},
  year={1996},
  volume={6},
  pages={15}
}

@article{Driver16,
  title={First-principles equation of state calculations of warm dense nitrogen},
  author={K.P. Driver and B. Militzer},
  journal={Phys. Rev. B},
  year={2016},
  volume={93},
  pages={064101}
}

@article{Militzer2021,
  title={First-principles equation of state database for warm dense matter computation},
  author={B. Militzer and F. Gonz\'{a}lez-Cataldo and S. Zhang and K. P. Driver and F. Soubiran},
  journal={Phys. Rev. B},
  year={2021},
  volume={103},
  pages={013203}
}

@article{Wu21,
  title={High-pressure phase diagram of beryllium from ab initio free-energy calculations},
  author={J. Wu and F. Gonz\'{a}lez-Cataldo and B. Militzer},
  journal={Phys. Rev. B},
  year={2021},
  volume={104},
  pages={014103}
}

@article{Karasiev16,
  title={Importance of finite-temperature exchange correlation for warm dense matter calculations},
  author={V.V. Karasiev and L. Calderin and S.B. Trickey},
  journal={Phys. Rev. E},
  year={2016},
  volume={93},
  pages={063207}
}

@article{Karasiev22,
  title={First-principles study of L-shell iron and chromium opacity at stellar interior temperatures},
  author={V.V. Karasiev and S.X. Hu and N.R. Shaffer and G. Miloshevsky},
  journal={Phys. Rev. E},
  year={2022},
  volume={106},
  pages={065202}
}

@article{SXHu11,
  title={First-principles equation-of-state table of deuterium for inertial confinement fusion applications},
  author={S.X. Hu and B. Militzer and V.N. Goncharov and S. Skupsky},
  journal={Phys. Rev. B},
  year={2011},
  volume={84},
  pages={224109}
}

@article{Ding18,
  title={Ab Initio Studies on the Stopping Power of Warm Dense Matter with Time-Dependent Orbital-Free Density Functional Theory},
  author={Y.H. Ding and A.J. White and S.X. Hu and O. Certik and L.A. Collins},
  journal={Phys. Rev. Lett.},
  year={2018},
  volume={121},
  pages={145001}
}

@article{Bonitz2024,
  title={First principles simulations of dense hydrogen},
  author={M. Bonitz and J. Vorberger and M. Bethkenhagen and M. P. Bohme and D. M. Ceperley and et al.},
  journal={Physics of Plasmas},
  year={2024},
  volume={31},
  pages={110501}
}

@article{Vinko12,
  title={Creation and diagnosis of a solid-density plasma with an X-ray free-electron laser},
  author={S. M. Vinko and O. Ciricosta and B. I. Cho and K. Engelhorn and H.-K. Chung and et al.},
  journal={Nature},
  year={2012},
  volume={482},
  pages={59}
}

@article{Kraus25,
  title={Warm dense matter studies with X-ray free-electron lasers},
  author={D. Kraus and T. R. Preston and U. Zastrau},
  journal={Nature Reviews Physics},
  year={2025},
  volume={8},
  pages={27}
}

@article{Bykova16,
  title={Structural complexity of simple $\text{Fe}_2\text{O}_3$ at high pressures and temperatures},
  author={E. Bykova and L. Dubrovinsky and N. Dubrovinskaia and M. Bykov and C. McCammon and et al.},
  journal={Nat. Commun.},
  year={2016},
  volume={7},
  pages={10661}
}

@article{Lavina11,
  title={Discovery of the recoverable high-pressure iron oxide $\text{Fe}_4\text{O}_5$},
  author={B. Lavina and P. Dera and E. Kim and Y. Meng and R. T. Downs and et al.},
  journal={PNAS},
  year={2011},
  volume={108},
  pages={17281}
}

@book{Cornell2003,
    author = {R. Cornell and U. Schwertmann},
    title = {The Iron Oxides: Structure, Properties, Reactions, Occurrences, and Uses},
    publisher = {Weinheim, Germany: Wiley-VCH},
    year = {2003}
}

@article{Ono2005,
  title={In situ x-ray observation of phase transformation in $\text{Fe}_2\text{O}_3$ at high pressures and high temper-
atures},
  author={S. Ono and Y. Ohishi},
  journal={J. Phys. Chem. Solids},
  year={2005},
  volume={66},
  pages={1714}
}

@article{Ono2004,
  title={High-pressure phase transition of hematite, $\text{Fe}_2\text{O}_3$},
  author={S. Ono and T. Kikegawa and Y. Ohishi},
  journal={J. Phys. Chem. Solids},
  year={2004},
  volume={65},
  pages={1527}
}

@article{Pasternak99,
  title={Breakdown of the Mott-Hubbard state in $\text{Fe}_2\text{O}_3$: A first-order insulator-metal transition with collapse of magnetism at 50 GPa},
  author={M. P. Pasternak and G. K. Rozenberg and G. Y. Machavariani and O. Naaman and R. D. Taylor and et al.},
  journal={Phys. Rev. Lett.},
  year={1999},
  volume={82},
  pages={4663}
}

@article{Badro02,
  title={Nature of the high-pressure transition in $\text{Fe}_2\text{O}_3$ hematite},
  author={J. Badro and G. Fiquet and V. V. Struzhkin and M. Somayazulu and H. K. Mao and et al.},
  journal={Phys. Rev. Lett.},
  year={2002},
  volume={89},
  pages={205504}
}

@article{Sanson16,
  title={Local structure and spin transition in $\text{Fe}_2\text{O}_3$ hematite at high pressure},
  author={A. Sanson and I. Kantor and V. Cerantola and T. Irifune and A. Carnera and et al.},
  journal={Phys. Rev. B},
  year={2016},
  volume={94},
  pages={014112}
}

@article{Greenberg18,
  title={Pressure-induced site-selective Mott insulator-metal transition in $\text{Fe}_2\text{O}_3$},
  author={E. Greenberg and I. Leonov and S. Layek and Z. Konopkova and M. P. Pasternak and et al.},
  journal={Phys. Rev. X},
  year={2018},
  volume={8},
  pages={031059}
}

@article{Celin26,
  title={Structural evolution of iron oxides melts at Earth’s outer-core pressures},
  author={C. Cr\'{e}pisson and M. Fitzgerald and D. Peake and P. G. Heighway and T. Stevens and et al.},
  journal={Nat Commun},
  year={2026},
  volume={},
  pages={},
  note={doi.org/10.1038/s41467-026-75204-4}
}

@article{Bussi20,
  title={Using metadynamics to explore complex free-energy landscapes},
  author={G. Bussi and A. Laio},
  journal={Nature Reviews Physics},
  year={2020},
  volume={2},
  pages={200}
}

@misc{github,
  author = {S. Azadi},
  title = {GitHub},
  url = {https://github.com/arshamsam}
}

\end{document}